\documentstyle[prl,aps,multicol,psfig]{revtex}
\begin{document}
\draft

\title{Nonlinear resonant tunneling in systems coupled to quantum reservoirs}
\author{Carlo Presilla$^a$ and Johannes Sj\"ostrand$^b$}
\address{
$^a$Dipartimento di Fisica, Universit\`a di Roma ``La Sapienza,''
Piazzale A. Moro 2, 00185 Roma, Italy \\
$^b$Centre de Math\'ematique, Ecole Polytechnique,
F-91128 Palaiseau Cedex, France and U.A. 169, C.N.R.S.}
\date{to be published in Phys. Rev. B}
\maketitle
\begin{abstract}
An adiabatic approximation in terms of instantaneous resonances is 
developed to study the steady-state and time-dependent transport of 
interacting electrons in biased resonant tunneling heterostructures.
The resulting model consists of quantum reservoirs coupled to regions 
where the system is described by nonlinear ordinary differential equations
and has a general conceptual interest.
\end{abstract}
\begin{multicols}{2}\narrowtext

The mathematical method recently proposed in \cite{JPS} provides 
a significant advance in the solution of time-dependent scattering
problems for Schr\"odinger equations with nonlinearities concentrated
near the resonances of the corresponding potential.
The method consists in the separation of the original system in two coupled
subsystems through the splitting of the wavefunction in two components.
In this way, we can separately study the simple problem of a reservoir-like
subsystem having only extended states and couple its solution to the other
subsystem having resonance states.   
The solution of this second problem is then simplified by an adiabatic 
approximation in terms of instantaneous resonances.

The situation investigated in \cite{JPS}
depicts ballistic transport in a double barrier heterostructure, i.e.,
the scattering of a wave-packet on a double barrier potential.
The nonlinearity, concentrated in the well between the barriers, is
due to the interaction of the electrons represented by the wave-packet
(mean-field).
Here, we generalize the approach \cite{JPS} to the case of biased
heterostructures where a band of scattering states are to be considered.
For these systems, hysteresis in the current-voltage 
characteristics has been observed \cite{GTC} and recognized
as a consequence of the mutual interaction of the electrons trapped
in the resonance \cite{GTC,ALL}.
We show that the approach \cite{JPS} allows us to quantitatively 
reproduce experimental results like \cite{ZGTC} and 
predict new time-dependent properties.
Although illustrated in the case of heterostructures, these results are
very general and have applications in fields like the theory of electric
systems \cite{WS} and nonlinear optics \cite{BONIFACIO,LUGIATO}.

Let us consider a heterostructure whose conduction band edge profile 
consists of two barriers of height $V_0$ located in $[a,b]$ and 
$[c,d]$ with $a<b<c<d$ along the growth direction $x$.
Translational invariance is assumed in the plane parallel to the junctions.
Suppose that a bias energy $\Delta V$ is applied between the emitter 
($x<a$) and collector ($x>d$) regions uniformly doped with net donor 
concentration $n_D$.  
At thermal equilibrium with temperature $T$, transport is due to 
a band of scattering states with Fermi energy 
$E_F= ( 3 \pi^2 n_D )^{2/3}$
(we use everywhere effective atomic units $\hbar=2m^*=1$ and 
$e^2/\varepsilon=2 a_B^{-1}$, where $m^*$ is the electron effective mass, 
$\varepsilon$ the dielectric constant and 
$a_B= \hbar^2 \varepsilon/(m^* e^2)$ the effective Bohr radius).
Due to the translational invariance in the plane parallel to the junctions,
the single-electron scattering state at energy $E$ along the $x$ direction
is described by the one-dimensional Schr\"odinger equation
\begin{equation}
\left[ -i \partial_t - \partial_x^2 + V_{\text{cb}}(x) + U(\phi,x) \right] 
\phi(x,t,E) = 0
\label{1DNSE}
\end{equation}
where $V_{\text{cb}}(x)$ is the step-like conduction band edge profile and 
$U(\phi,x)$ takes into account the applied bias and, at Hartree level, 
the electron-electron interaction.
Assuming ideal metallic behavior in the emitter and collector regions,
i.e., neglecting the formation of accumulation and depletion layers,
$U(\phi,x)$ can be obtained as solution of the Poisson equation
\begin{equation}
\partial_x^2 U(\phi,x) = - 8 \pi a_B^{-1} \int dE~g(E)~|\phi(x,t,E)|^2
\label{POISSON}
\end{equation}
with boundary conditions $U(\phi,a)=0$, $U(\phi,d)=-\Delta V$. 
The parallel degrees of freedom are considered through
\begin{equation}
g(E) = {\Theta(E) \over 2 \pi} 
\left[ k_B T \ln \left( 1 + e^{E-E_F \over k_B T} 
\right) + E_F - E \right], 
\label{G}
\end{equation}
where the Heaviside function $\Theta(E)$ limits the integration in
(\ref{POISSON}) to energies above the bottom of the emitter conduction
band, $E=0$, as correct for $E_F \ll \Delta V$.

In general, the solution of (\ref{POISSON}) can not be handled analytically.
We will suppose that, 
due to the accumulation of electrons in the well with sheet density 
\begin{equation}
s(\phi) = \int dE~g(E)~\int_{(a+b)/2}^{(c+d)/2} dx~|\phi(x,t,E)|^2,
\label{S}
\end{equation}
ideal metallic behavior in the well $[b,c]$ and ideal insulating behavior
in the barriers $[a,b]$ and $[c,d]$ hold. 
Then the total potential $V_{\text{cb}}+U$ in (\ref{1DNSE}) is better 
rewritten as $V+W$ where
\begin{eqnarray}
V(x) &=& [V_0 - \Delta V (x-a)/\ell]~1_{[a,b]}(x) 
- \Delta V (b-a)/\ell~ \nonumber \\
&& \times 1_{[b,c]}(x)
+ [V_0 - \Delta V (b-a + x-c)/\ell]~  \nonumber \\
&& \times 1_{[c,d]}(x) - \Delta V~ 1_{[d,+\infty[}(x)
\label{V}
\end{eqnarray}
gives the band profile modified by the external bias and
\begin{eqnarray}
W(s,x) &=& 8 \pi a_B^{-1}~s(\phi)~ \bigl\{  
 (x-a)(d-c)/\ell ~1_{[a,b]}(x)  \nonumber \\
&& +(b-a)(d-c)/\ell ~1_{[b,c]}(x) \nonumber \\ 
&& +(b-a)(d-x)/\ell ~1_{[c,d]}(x) \bigr\}
\label{W}
\end{eqnarray}
depends on the wavefunction $\phi$ through the sheet density 
of electrons in the well $s(\phi)$.
Here $\ell=b-a + d-c$ and $1_{[x_0,x_1]}(x)$ is 1 if $x \in [x_0,x_1]$
and 0 otherwise.

Following \cite{JPS}, we will solve (\ref{1DNSE}) with the potential
(\ref{V}-\ref{W}) in two steps.
Let $V_{\text{fill}}(x) = V(x) + V_0 1_{[b,c]}(x)$ be the potential obtained
by filling the well $[b,c]$. 
First we solve 
\begin{equation}
\left[ -i \partial_t - \partial_x^2 + V_{\text{fill}}(x) + W(s,x)  
\right] \tilde{\mu}(x,t,E) = 0
\label{MUSE}
\end{equation}
and then we look for $\phi$ in the form $\phi=\tilde{\mu}+\tilde{\nu}$ 
where $\tilde{\nu}(x,t,E)$ should solve
\begin{eqnarray}
\left[ -i \partial_t - \partial_x^2 + V(x) + W(s,x)  \right] 
\tilde{\nu} = V_0~1_{[b,c]}(x) \tilde{\mu}.
\label{NUSE}
\end{eqnarray}
Substituting (\ref{1DNSE}) with (\ref{MUSE}-\ref{NUSE}) corresponds 
to decomposing the original system in quantum reservoirs coherently
coupled to a small subsystem.
The wave function $\tilde{\mu}$ describes an electron at energy $E$ 
which is delocalized in the emitter and collector regions (reservoirs) and
has an exponentially small probability to be found in the forbidden region
$[a,d]$.
The wave function $\tilde{\nu}$ describes the same electron in the double
barrier region and is driven by the value of $\tilde{\mu}$ in the well
$[b,c]$.
Due to the quasi-localization of $\tilde{\nu}$ in $[b,c]$,
we have $\phi \simeq \tilde{\mu}$ in the reservoirs and
$\phi \simeq \tilde{\nu}$ in the well, with an error which is exponentially
small in the limit of wide barriers \cite{JPS}.

Equation (\ref{MUSE}) can be solved by evaluating the instantaneous
eigenstates of the potential $V_{\text{fill}}+W$.
We put $\tilde{\mu}(x,t,E)=\exp(-iEt)\mu(x,t,E)$ and suppose
that $\Delta V$ and $s$ are slowly varying functions of time so that
also $\mu(x,t,E)$ is slowly varying in time. 
In the emitter region $x<a$ we take $\mu(x,t,E)=\mu(x,E)$ as the sum of
a left- and right-going plane wave at energy $E$ 
and propagate this expression to the adjacent regions by requiring 
$\mu$ to be of class $C^1$.
For wide barriers we can use a WKB expansion for the potential 
$V_{\text{fill}}+W$ and explicitly evaluate $\mu$ in the region 
$[b,c]$ which is of interest for solving (\ref{NUSE}).

Equation (\ref{NUSE}) can be simplified by developing $\tilde{\nu}$ 
into the instantaneous eigenstates of the potential $V+W$ and keeping
only the contributions from the discrete resonant states \cite{JPS}.  
In the case of a single resonant state we put 
$\tilde{\nu}(x,t,E)=\exp(-iEt) z(t,E) e(s,x)$ 
where $e(s,x)$ is the (ground) resonant state of the potential $V+W$
with complex eigenvalue $\lambda(s)=E_R(s)-i\Gamma(s)/2$
\begin{equation}
\left[ -\lambda(s) - \partial_x^2 + V(x) + W(s,x)  \right] e(s,x) = 0.
\label{ES}
\end{equation}
The eigenfunction $e(s,x)$ is of class $L^2$ on the contour 
$\gamma \equiv \left( e^{i\theta} ]-\infty,0]+a \right) 
\bigcup ~[a,d] \bigcup ~\left( d+e^{i\theta}[0,+\infty[ \right)$ 
for $\theta$ conveniently chosen \cite{AC} and satisfies
$\int_\gamma dx~e(s,x)^2 = 1$,
$\int_\gamma dx~e(s,x)~\partial_s e(s,x) = 0$. 
Multiplying (\ref{NUSE}) with $e$ and integrating over $\gamma$, we get
\begin{equation}
\partial_t z(t,E) = 
i \left[ E - \lambda(s) \right] z(t,E) + {\cal B}(s,E)
\label{ZDOT}
\end{equation}
with the driving term given by 
${\cal B}(s,E) = i V_0 \int_b^c dx~ \mu(x,t,E) e(s,x)$
and the sheet density (\ref{S}) reduced, with small error, to
\begin{equation}
s = \int dE~g(E)~|z(t,E)|^2 \equiv \|z(t)\|^2.
\label{NZ2}
\end{equation}
Explicit expressions of $\lambda(s)$ and $e(s,x)$ can be found within 
the same WKB approximation used for evaluating $\mu$  \cite{PS}.
For later use we note that $E_R(s) = E_R^0 + \eta s$, 
where $E_R^0=E_0 - \Delta V (b-a)/\ell$, 
$E_0$ being the (ground) eigenstate of the potential
$V_0 \left[ 1_{]-\infty,b]}(x) + 1_{[c,+\infty[}(x) \right]$
and $\eta= 8\pi a_B^{-1} (b-a)(d-c)/\ell = 
e^2/(C_{\text{e}}+C_{\text{c}})$, $C_{\text{e}}$ and $C_{\text{c}}$ 
being the emitter and collector capacitance per unit area, respectively.
Moreover, $\Gamma(s) = \Gamma_{\text{e}}(s)+\Gamma_{\text{c}}(s)$,
$\Gamma_{\text{e}}$ and $\Gamma_{\text{c}}$ being the contributions to the
resonance width given by the emitter and collector barriers, respectively.

The original problem (\ref{1DNSE}) is reduced to solving the system 
(\ref{ZDOT}) with the condition (\ref{NZ2}). 
Let us first consider the stationary solutions 
\begin{equation}
z(E) = { {\cal B}(s,E) \over 
- {\Gamma(s)/2} + i \left[ E - E_R(s) \right] }.
\label{ZFIX}
\end{equation}
Equation (\ref{NZ2}) gives a self-consistency condition for 
$s=\|z\|^2$. 
Assuming that $|{\cal B}(s,E)|^2$ is a smooth function of 
$E$ and $\Gamma(s) \ll E_F$, a Dirac $\delta$ approximation can be used to get
\begin{equation}
s = f(s) \equiv 2\pi~ g(E_R(s))~
\left| {\cal B}\left( s,E_R(s) \right) \right|^2~ \Gamma(s)^{-1}.
\label{SGA}
\end{equation}
The function $f(s)$ vanishes everywhere except for
$0 \leq E_R(s) \lesssim E_F$ where, for $E_F \ll V_0$ we have 
$2\pi \left| {\cal B}(s,E_R(s)) \right|^2\simeq\Gamma_{\text{e}}(s)$.
Equation (\ref{SGA}) is then equivalent to
$s \Gamma(s) = g(E_R(s)) \Gamma_{\text{e}}(s)$ which has a simple
interpretation in terms of charge conservation.
In the steady-state, the current of electrons injected from the emitter
into the well, $g(E_R(s)) \Gamma_{\text{e}}(s)$, equilibrates
the escaping current, $s \Gamma(s)$.
The latter current increases with increasing the sheet density of
electrons in the well, $s$, while the former vanishes at
both $E_R(s) =0$ (square root singularity) and $E_R(s) \simeq E_F$.
Therefore, Eq. (\ref{SGA}) has only one solution for $E_R^0 \geq 0$
and may have three for $E_R^0 < 0$.
For $E_R^0 \gtrsim E_F$ the unique solution vanishes and
for $E_R^0 < 0$ the couple of nonvanishing solutions is to be searched
in the interval $-E_R^0/\eta \leq s \lesssim (-E_R^0 + E_F)/\eta$.

In terms of applied bias, multiple solutions of (\ref{SGA}) can be 
obtained for $\Delta V > E_0 \ell/(b-a)$.
The range of $\Delta V$ values for which three solutions exist
depends on the amplitude of the function $f$.
If the emitter barrier is more opaque than the collector one,
$f$ is suppressed by the factor
$\Gamma_{\text{e}} / \Gamma_{\text{c}} \ll 1$
and we always have only one solution.

The solutions of (\ref{SGA}) can be characterized in terms
of stability.
This is particularly important in view of a comparison with steady-state
experiments where only stable 
\begin{figure}
\centerline{\hbox{\psfig{figure=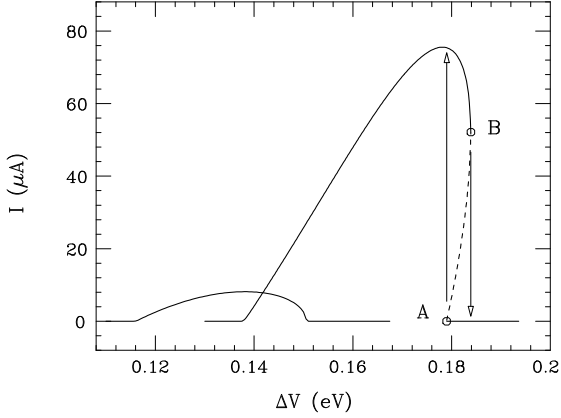,width=8.5cm,angle=90}}}
\caption{
Theoretical steady-state current-voltage characteristic for the GaAs-AlGaAs
heterostructure experimentally investigated in \protect\cite{ZGTC}
under forward (right-most curve) and reverse (left-most curve) bias.
In the reverse bias case we permuted the barriers instead of making
$\Delta V$ (and $I$) negative.
In the forward bias case the dashed line is an unstable steady-state solution 
and arrows indicate the transition expected at the bistability thresholds 
A and B by decreasing or increasing the bias, respectively.
The relevant parameters are
$n_D=2 \times 10^{17}$ cm$^{-3}$, $T=1$ K, $A=2 \times 10^{-5}$ cm$^2$,
$m^*=0.067~m$, where $m$ is the free electron mass, $\varepsilon=11.44$,
$V_0=0.34$ eV, $b-a=9.0$ nm, $c-b=5.6$ nm, and $d-c=10.7$ nm
\protect\cite{EQASYM}.}
\label{FIG1}\end{figure}\noindent
solutions are measured.
By studying the eigenvalues of the linearization of the vector field
defined by the r.h.s. of (\ref{ZDOT}) one can demonstrate that
a solution of (\ref{SGA}) is stable (unstable) 
when $\partial_s f(s) < 1$ ($> 1$) \cite{PS}. 
The trivial solution $s=0$, when it exists, is, therefore, a stable
one. 
When three solutions exist, two of them, the largest and the smallest one,
are stable while the intermediate one is unstable.

Considerations analogous to those made for $s$ hold for the steady-state
collector current $I/e =  A s ~ \Gamma_c(s)$, proportional to the number
of electrons in the well, $As$, $A$ being the transverse area of the
heterostructure, and to collector escape rate $\Gamma_c(s)$.
In Fig. 1 we show $I(\Delta V)$ evaluated for the asymmetric double-barrier
heterostructure experimentally investigated in \cite{ZGTC}.
In agreement with the above discussion and with the experimental findings,
no multiple solutions are obtained in the left-most curve (reverse bias case)
of Fig. 1 when the emitter barrier is wider than the collector one. 
On the other hand, a bistability region extending between points A and B 
is observed in the forward bias case.

The above results can be generalized to include the effect of inelastic
processes if we change $\Gamma \to \Gamma + \Gamma_i$ in Eq. (\ref{ZFIX}),
where $\Gamma_i=\Gamma_{ie}+\Gamma_{ic}$ is the total width
representing collector and emitter inelastic decay channels \cite{STONELEE}.
The collector current becomes $I/e =  A s ~ (\Gamma_c(s)+\Gamma_{ic})$.
However, if $\Gamma + \Gamma_i \ll E_F$, Eq. (\ref{SGA}) still holds
so that for $\Gamma_c \simeq \Gamma$ ($\Gamma_{ic} \simeq \Gamma_{i}$),
as in the case of Fig. 1, the value of $I$ is independent of the ratio
$\Gamma_i/\Gamma$.
\begin{figure}
\centerline{\hbox{\psfig{figure=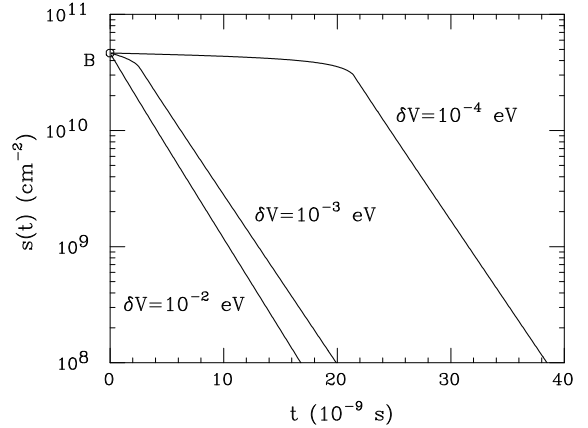,width=8.5cm,angle=90}}}
\caption{Sheet density of electrons in the well  $s(t)$ after an 
instantaneous increase $\delta V$ of the bias from point $B$ of Fig. 1.}
\label{FIG2}\end{figure}\noindent

Now we turn to the time-dependent transport properties.
According to (\ref{ZDOT}) and (\ref{NZ2}), we have
\begin{eqnarray}
\partial_t s(t) = - \Gamma(s(t))~s(t)
+ 2 \text{Re} \langle z | {\cal B} \rangle 
\label{FNZ2DOT}
\end{eqnarray}
where $\langle u | v \rangle \equiv \int dE~g(E)~\overline{u(E)} v(E)$.
The last term in (\ref{FNZ2DOT}) can be expressed in terms of 
$s(t)$ by using the formal solution of (\ref{ZDOT})
\begin{eqnarray}
&&z(t,E) = e^{ \int_0^t dt'  
\left[ - {1\over2} \Gamma(s(t'))
+ i \left( E-E_R(s(t')) \right) \right] } ~z(0,E) +
\nonumber \\
&& \int_0^t dt' e^{ \int_{t'}^t dt''  
\left[ - {1\over2} \Gamma(s(t''))
+ i \left( E-E_R(s(t'')) \right) \right] }
{\cal B}\left( s(t'),E \right).
\label{ZSOL}
\end{eqnarray}
The first term in (\ref{ZSOL}) vanishes exponentially and
can be neglected after a time $t \gg 2/\Gamma$.
In the second term, an analogous exponential factor selects the 
contributions for $t-t' \lesssim 2/\Gamma$ as the dominant ones so that 
the lower integration bound can be safely changed into $-\infty$
for $t \gg 2/\Gamma$.
In this case, $2 \text{Re} \langle z | {\cal B} \rangle$ can be approximated 
with 
\begin{eqnarray}
&&2 \text{Re} \int_{-\infty}^{~t} dt'~
e^{ \int_{t'}^t dt'' \left[ - {1\over2} \Gamma(s(t''))  
+ i E_R(s(t'')) \right] } {\cal F}(g|{\cal B}|^2) (t-t')
\nonumber \\ &&\simeq 
2 \text{Re} \int_{-\infty}^{~t} dt'
~e^{ i E_R(s(t)) (t-t') } {\cal F}(g|{\cal B}|^2) (t-t')
\nonumber \\ &&= 
2 \pi~ g \left( E_R(s(t)) \right)~
\left| {\cal B} \left( s(t), E_R(s(t))  \right) \right|^2,
\label{REZB}
\end{eqnarray}
where ${\cal F}(g|{\cal B}|^2)$ is the Fourier transform obtained by
performing the energy integral in the scalar product and the
approximation in the third line is valid for $\Gamma$
and $\partial_t s(t)$ small.
With this result Eq. (\ref{FNZ2DOT}) becomes
\begin{equation}
\partial_t s(t) = - \Gamma(s(t))~ [ s(t) - f (s(t)) ].
\label{NZ2DOT}
\end{equation}
As Eq. (\ref{SGA}), this is a conservation law for the charge trapped
in the well.
The steady-state solutions, $\partial_t s = 0$, of (\ref{NZ2DOT}) 
coincide with those defined by (\ref{SGA}) and also
their 
\begin{figure}   
\centerline{\hbox{\psfig{figure=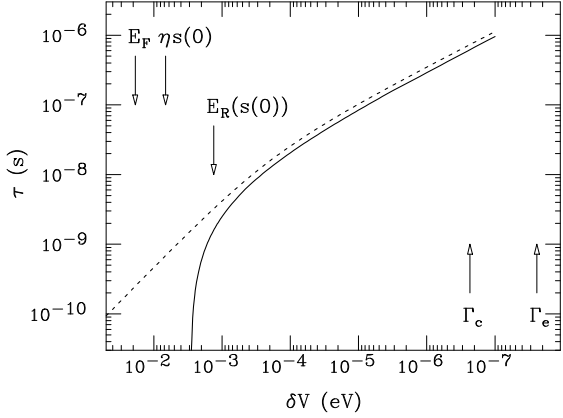,width=8.5cm,angle=90}}}
\caption{
Survival time of the quasi-stationary solutions 
of Fig. 2 {\it versus} $\delta V$ (solid line).
The dashed line is Eq. (\protect{\ref{TS}}).
Arrows indicate the relevant energy scales.}
\label{FIG3}\end{figure}\noindent
attractive nature agrees with the stability condition discussed above.

Close to a bistability threshold, the dynamics of certain nonlinear optical
systems has been shown to be characterized by a quasi-stationary transient 
followed by a fast evolution \cite{BONIFACIO,LUGIATO}.
As we will discuss in a moment, this behavior is typical of any
rate equation of the form $\partial_t s = h(s,\delta V)$ where $\delta V$
is the control parameter of a bistability threshold.
An example for our heterostructure is shown in Fig. 2.
At time $t=0$ the system is in the threshold stable steady-state $B$ of
Fig. 1 when the bias is instantaneously increased by an amount $\delta V$.
For $\delta V$ smaller than a critical value, we observe a 
quasi-stationary $s(t)$ which decreases linearly at small times
and, after a time $\tau$, vanishes in a nearly exponential way.
The border between these two regimes is given by the condition 
$E_R(s(\tau))=0$.
Indeed, when the resonant energy $E_R(s)$ becomes smaller than the emitter 
band edge the filling current, $\Gamma(s) f(s)$, vanishes and (\ref{NZ2DOT})
has solution $s(t) \propto \exp[-\int^t dt' \Gamma(s(t'))]$.
The nearly exponential decay is established from the beginning if
$\delta V \gtrsim E_R(s(0))$.
For $\delta V < E_R(s(0))$ and $t \leq \tau$, $s(t)$ is in a quasi-stationary 
regime which is characterized only by the fact that the starting point,
$s(0)$, is a threshold stable steady-state.
Indeed, for $s(t) -s(0)$ and $\delta V$ small in this case we must have 
$h(s,\delta V) \simeq -q [s(t)-s(0)]^2 -p \delta V$ with $q,p>0$,
independently of $h(s,\delta V)$.
Integrating, we get $s(t)=s(0)-\sqrt{\delta V p/q}
\tan( \sqrt{qp\delta V} t)$ which has linear behavior for small $t$. 
In the case of Eq. (\ref{NZ2DOT}), the condition $E_R(s(\tau))=0$ and
the approximate evaluation of $q$ and $p$ at $T=0$ K give 
\begin{equation}
\tau \simeq {4 E_R(s(0)) \over \eta s(0) \Gamma_{\text{c}} } 
\sqrt{E_R(s(0)) \over \delta V}~ \arctan
\left( \sqrt{E_R(s(0)) \over 4 \delta V} \right)
\label{TS}
\end{equation}
where $E_R(s(0)) \simeq E_F \Gamma_{\text{c}} /(2 \Gamma)
[1+2\pi\Gamma /(\eta \Gamma_e)]^{-1}$  and
$s(0) \simeq [E_F -E_R(s(0))] \Gamma_{\text{e}} /(2 \pi \Gamma)$. 
When $\delta V \ll E_R(s(0))$, we have $\tau \sim  \delta V^{-1/2}$
as shown in Fig. 3 where we compare (\ref{TS}) with $\tau$
obtained by numerically integrating (\ref{NZ2DOT}) \cite{VALIDITY}.
For $E_R(s(0)) \ll E_F$, the temperature dependence of (\ref{TS}) is easily
obtained by substituting the Fermi energy with the effective
value $\widetilde{E}_F = E_F +k_BT \ln[1+\exp(-E_F/k_BT)]$.
The survival time increases by increasing $T$ and for
$\delta V \ll E_R(s(0))$ we have $\tau \sim \widetilde{E}_F^{1/2}$.

The phenomenon discussed above may be exploited for device applications
like those suggested for optical systems \cite{BONIFACIO}.
More complex time-dependent features are expected in heterostructures 
with many resonances \cite{PS} or in superlattices \cite{KGPPWS}.

It is a pleasure to thank G. Jona-Lasinio for stimulating discussions, 
as well as G. Perelman, who independently observed how to get 
(\ref{NZ2DOT}) from (\ref{FNZ2DOT}).
We are indebted to F. Capasso and to an anonymous referee for suggesting 
the connection with optical nonlinear phenomena and to L. A. Lugiato 
for pointing out to us \cite{LUGIATO}.

\end{multicols}

\begin{references}
\bibitem{EQASYM} We simulate the asymmetry of the experimentally 
investigated heterostructure having equal-width different-height barriers
by choosing equal-height different-width barriers with equivalent opacity.
\bibitem{VALIDITY}
Equation (\protect\ref{NZ2DOT}) is an approximation of
(\protect\ref{FNZ2DOT}).
This approximation, however, is fully appropriate to study the 
survival time in the interesting region $\delta V < E_R(s(0))$
where $\tau \gg \Gamma^{-1} \sim 10^{-9}$ s.
This is confirmed by a comparison with the exact results 
obtained by numerically solving (\protect\ref{ZDOT}) on a proper 
time-energy lattice \protect\cite{PS}.
\end{references}
\end{document}